\documentclass{optica-article}

\journal{opticajournal} 

\articletype{Research Article}

\usepackage{graphicx}

\usepackage{IEEEtrantools}

\usepackage{enumitem}

\usepackage{mathtools}
\usepackage{hyperref}
\usepackage{amsfonts}
\usepackage[T1]{fontenc}

\usepackage{algorithm}
\usepackage{algpseudocode}

\DeclareGraphicsExtensions{.png,.pdf,.eps}

\newcommand{\figref}[1]{Figure~\ref{#1}}

\newcommand{\secref}[1]{Sec.~\ref{#1}}

\def \diracspacing {0.7pt}
\newcommand{\ket}[1]{| \hspace{\diracspacing} #1 \rangle} 

\begin{document}

\title{Greedy receiver for photon-efficient optical communication}

\author{Karol Łukanowski}

\address{Centre for Quantum Optical Technologies, Centre of New Technologies, University of Warsaw, Stefana Banacha 2c, 02-097 Warszawa, Poland}

\email{k.lukanowski@cent.uw.edu.pl} 


\begin{abstract*} 
In optical communication the transmitter encodes information into a set of light states defined by the modulation format, selected to accommodate specific channel conditions and to remain sufficiently distinguishable at the output. Various receiver architectures have been designed to improve the demodulation performance, ultimately limited by quantum theory. In this work I introduce a new receiver based on a locally optimal greedy algorithm and apply it to pulse position modulation. The receiver reduces the error probabilities of previously proposed strategies in all signal strength regimes and achieves results comparable with those obtained by numerical optimization of the detection process. In contrast, however, it is conceptually simple and therefore can be scaled to arbitrarily high modulation orders for which numerical methods become intractable. In the photon-starved regime characteristic of deep space optical communication, the greedy receiver approaches the quantum-optimal Helstrom bound on state discrimination error probability. In the regime of few-photon pulses, the error reduction offered over the other methods grows up to an order of magnitude.

\end{abstract*}


\section{Introduction}
%
Long distance communication between devices of modern technology is for the most part optical. Messages are transmitted by modulating an electromagnetic wave and propagating it through an optical channel, such as fiber or free space, upon which end the receiver strives to demodulate the incoming signal to read out the original message~\cite{Shannon1949,mackay2003,Banaszek2020}. Although research into the fundamental quantum description of light paves the way for novel non-classical light-state encodings of information that may prove advantageous in specialized applications---e.g., photon-number states in purely lossy conditions~\cite{Lukanowski2021} or squeezed light in strongly dephasing channels~\cite{Fanizza2021}---the modulation techniques currently in use still rely on essentially classical laser pulses~\cite{Banaszek2020}. These are represented in quantum theory by coherent states of light denoted $\ket{\alpha}$. The modulus and argument of the complex number $\alpha$ are then, respectively, the amplitude and the phase of the classical lightwave~\cite{Glauber1963,Sudarshan1963}.

Surprisingly, although coherent states are never mutually orthogonal and thus cannot be distinguished with zero probability of error by any physical measurement, they can still saturate the ultimate Holevo limit on achievable bitrate in optical information transmission~\cite{Giovannetti2004,Giovannetti2014,Holevo1973,Schumacher1997}. Such a communication protocol would, however, require a receiver able to perform collective measurements on multiple incoming states, the optical implementation of which remains challenging~\cite{Guha2011theo2,Sasaki1998}. In practice, several modulation formats based on coherent states have been established and the research into practical receiver architectures that improve demodulation performance continues~\cite{Bilkis2020}. 


In this work I introduce a new coherent state receiver that adapts its behaviour during signal reception. Crucially, it does so according to a so-called \emph{greedy} algorithm belonging to a class of methods in computer science that make locally optimal choices in the hope that they will lead to a globally optimal solution~\cite{Cormen2022}. One classic illustration of such algorithms involves the knapsack problem in which, given a set of objects with weights and values, one (perhaps a thief in a jewelry store) is tasked with filling the knapsack up to some weight while maximizing the total value of the items taken. A greedy approach would be to sort the items by value-to-weight ratio and pack them one by one until the maximal weight is reached. If the choice is always to either take an item or leave it, this usually produces suboptimal solutions. Packing certain items early on limits the ability to later pack other ones, that although possess a lower ratio, could ultimately yield a higher total value and still fit under the weight limit. However, the greedy solution is easily computable and oftentimes \emph{good enough}, in contrast to globally optimal methods (that do not do what is best at the moment, but rather consider far-reaching consequences of their choices) which inevitably consume more resources and time. Furthermore, in some cases greedy turns out to be provably optimal---such as the fractional knapsack problem, in which one is allowed to take fractions of items (for instance, if the thief has the option to cut the precious jewels). This work studies the greedy approach to optical demodulation on the exemplary pulse position modulation format and shows it to be very effective.

\section{Pulse position demodulation}

\begin{figure*}[t]
\makebox[\textwidth][c]{\includegraphics[width=1.3\textwidth]{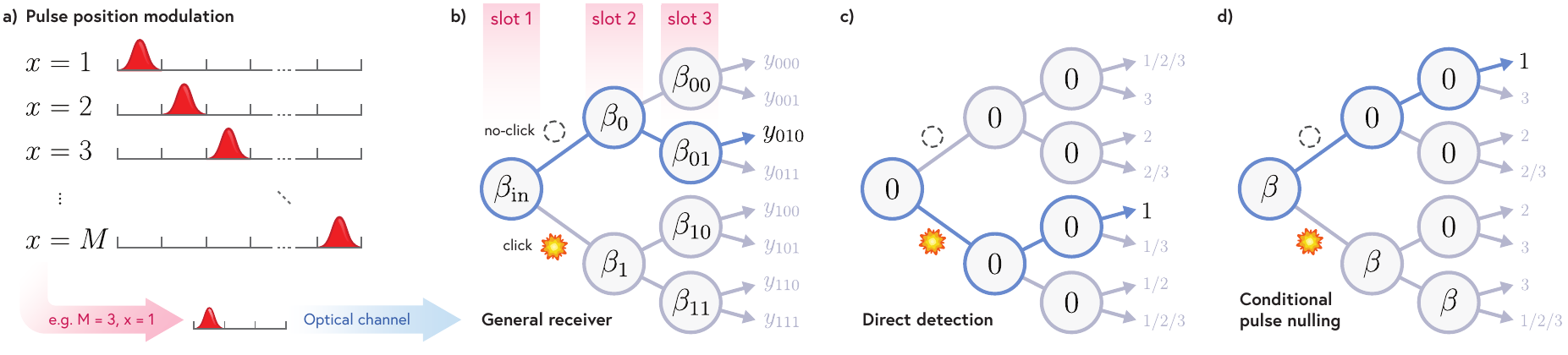}
}
\caption{\textbf{a) Pulse position modulation.} In one channel use of $M$-PPM, the transmitter produces a symbol $x \in \{1, \ldots, M\}$ by dividing a time frame into $M$ slots and preparing a pulse in one of them, leaving the other slots empty. The symbol is then transmitted through an optical communication channel to the receiver, whose shutter is synchronized with the transmitter so that the detection begins with the arrival of the frame. The aim of the receiver is to output an estimate $y$ of the pulse position in the received frame that agrees with the actual symbol $x$ being transmitted. An exemplary transmission of $x=1$ in 3-PPM is depicted. \textbf{b)---d) PPM demodulation algorithms with displacements.} Standard PPM demodulation strategies measure the incoming signal in each slot one by one with ``on-off'' direct detection resulting in a ``click'' if photons are detected or a ``no-click'' otherwise. The receiver outputs some final pulse position estimate $y$ by following a decision tree with branches corresponding to the binary measurement outcomes. Displacement receivers additionally shift the slot amplitude before direct detection by some amount $\beta$ prescribed in the tree nodes. The trees corresponding to direct detection and conditional pulse nulling are depicted for 3-PPM and highlighted are some possible paths through the tree leading to the correct estimate $y=1$. }
\label{fig:fig_ppm_rec}
\end{figure*}

In severely power-limited conditions, such as deep-space optical communication, pulse position modulation (PPM) is the common modulation choice~\cite{Hemmati2006,Hemmati2011,Boroson2004}. This is in part due to straightforward signal preparation and high photon information efficiency approaching the quantum-optimal Holevo limit in the photon-starved regime
~\cite{Kochman2014,Jarzyna2015,Jarzyna2024}. In $M$-ary PPM or $M$-PPM, where we refer to $M$ as the PPM \emph{order}, sketched in~\figref{fig:fig_ppm_rec}a), the frame of chosen duration $T$ is divided into $M$ \emph{slots}. A pulse of light is prepared in one of them, denoted by a \emph{symbol} $x \in \{1, \ldots, M\}$. The position of the pulse within the frame carries then an information content of up to $\log_2 M$ bits~\cite{mackay2003}. To extract this information, the aim of the receiver is to identify the pulse position within the received window, or in other words, measure the arrival time of the pulse within a single frame and output an \emph{estimate} $y \in \{1, \ldots, M \}$ that agrees with $x$.

A simple way to assess receiver performance is to calculate the average probability $P_e$ of an incorrect identification of the input symbol, or, equivalently, the average probability of correct decision $P_c$. We have
\begin{equation}
\label{eq:prob_correct}
P_c = \sum_{x,y} p(x) p(y=x|x), \qquad P_e = 1 - P_c,
\end{equation}
where $p(y=x|x)$ is the probability of correctly identifying the symbol $x$ by the output estimate $y$, conditioned on $x$ being transmitted which happens with probability $p(x)$. Typically one sets all $p(x)$ equal, so that $P_c = \sum_{x,y} p(y=x|x) / M$. Thanks to the celebrated Helstrom theory~\cite{Helstrom1969}, it is possible to tightly lower-bound $P_e$, although only for highly symmetric state constellations~\cite{Cariolaro2010}. In PPM, for instance, if the received pulse slot indeed holds a pure coherent state $\ket{\alpha}$ and the other slots contain a perfect quantum vacuum $\ket{0}$, which corresponds to ideal noiseless transmission, the minimal error probability reads~\cite{Pozza2014}
\begin{IEEEeqnarray}{rCl}
\label{eq:Helstrom}
P_{e,M}^{\mathrm{H}} &=& \frac{M-1}{M^2} \! \left(\!\sqrt{1\!+\!(M-1) e^{-|\alpha|^2}} - \! \sqrt{1\! -\! e^{-|\alpha|^2}} \right)^{\!\!2}\!\!.
\IEEEeqnarraynumspace
\end{IEEEeqnarray}
This ultimate limit provides a benchmark to which the performance of any receiver can be compared, although for this comparison to be fair, similar noiseless or near-noiseless conditions should be assumed.

For more realistic scenarios in which the transmitted states are disturbed by loss and noise, no general quantum limit is known~\cite{Cariolaro2010}, but it is still customary and convenient to compare receivers based on their error probabilities~\cite{Pozza2014}. Note, however, that for the sake of simplicity, some nuance is lost. First of all, besides errors, receivers could also announce erasures (the ``I have no idea'' output) which can be beneficial for the bitrate~\cite{mackay2003}---while to minimize the error probability it is always better to guess the input symbol randomly in the event of an erasure, so that at least sometimes the output estimate will be correct. Still, the addition of erasures is straightforward for all the receivers that will be studied here. Second of all, the intuition ``the smaller the error probability, the higher the protocol bitrate'' comes with some caveats. From some point onward, decreasing the error probability does not increase the bitrate much if it is already close to the Shannon limit~\cite{mackay2003}. Nonetheless, smaller error probabilities always lead to improvements in performance, if not in the bitrate itself, then in error-correction coding complexity~\cite{CoverThomas,Chen2012} and finite-blocklength communication~\cite{Polanskiy2010}, such as that between Earth and a satellite that passes only briefly over the ground station. I shall therefore keep the error probability as a figure of merit.

\subsection{The direct detection receiver}

The basic approach to PPM demodulation is the direct detection (DD) receiver that relies on photon counting in each subsequent slot of the frame to estimate the pulse position. Typically the detection is performed by a single photon detector (SPD) operating in ``on-off'' mode, so that either no detections are observed in a slot (a ``no-click'') or at least one is (a ``click'')~\cite{Jarzyna2017,Zwolinski2018}. Most of the time one can expect that the pulse slot $x$ will produce a click and the other empty slots will result in no-clicks, leading to the correct estimate $y = x$. If a click is recorded in more than one slot due to stray light, background noise, or dark counts in the detector, the receiver outputs an estimate chosen randomly out of these slots. On the other hand, the stochastic nature of photodetection makes it possible to detect no photons in the frame at all, even in the pulse slot due to the nonzero vacuum contribution to the state. Then the receiver could announce an erasure, but under the error-only assumption it simply guesses randomly from all the $M$ slots. The DD error probability is given by (see the Supplementary Material for the derivation)
\begin{equation}
\label{eq:PeDD}
P_{e,M}^{\mathrm{DD}} = \frac{(M - p_0 q_0^{M-1}) \bar{q}_0 - \bar{p}_0 (1-q_0^M)}{M \bar{q}_0},
\end{equation}
where $p_0$ is the no-click probability in the pulse slot (${\bar{p}_0=1 - p_0}$) and $q_0$ the no-click probability in an empty slot ($\bar{q}_0 = 1 - q_0$). Some authors refer to~\eqref{eq:PeDD} as the \emph{standard quantum limit}~\cite{Chen2012}. \figref{fig:fig_ppm_rec}c) depicts an exemplary decision tree traversed by the DD receiver in the detection process.

\subsection{Displacement receivers and conditional pulse nulling}
A broader class of receivers introduced by Kennedy and Dolinar for a binary coherent state modulation~\cite{Kennedy1973,Dolinar1973} and later adapted by Dolinar to PPM~\cite{Dolinar1983} are \emph{displacement receivers} that displace in phase-space the complex amplitudes of incoming signals before performing direct detection on them. If the incoming state is a pure coherent state $\ket{\alpha}$, displacing by $\beta$ yields $\ket{\alpha - \beta}$. In practical models with different kinds of noise the mathematical description becomes more involved, but the intuition of displacing the amplitude in the slot holds. The goal of displacement is to modify the photodetection statistics in a way that can be exploited to reduce the receiver error probability. Such a strategy has proved useful in receivers designed for other modulation formats as well~\cite{Bondurant1993,Burenkov2021}.

Without loss of generality, let us denote by $q_\beta$ the probability of no clicks being recorded in an empty slot that was displaced by $\beta$ and the corresponding click probability by $\bar{q}_\beta := 1 - q_\beta$. Similarly, let $p_\beta$ be the no-click probability in the slot containing the pulse and displaced by $\beta$, and $\bar{p}_\beta := 1 - p_\beta$ the corresponding click probability. Poissonian photon statistics~\cite{Mandel1995} assumed in both the signal and displacement modes, as well as for the additive noise, result in 
\begin{equation}
\label{eq:poisson}
q_\beta = e^{- \beta^2 - N_b }, \quad p_\beta = e^{-\left(\alpha - \sqrt{1-\Delta} \beta\right)^2 - \Delta \beta^2 - N_b }.
\end{equation}
Here $\alpha$ is the pulse amplitude, $\beta$ is the displacement applied in the round, and $N_b$ is the average number of noise photons impinging on the detector in one slot. The parameter $\Delta$ represents mode mismatch between $\alpha$ and $\beta$---if positive, their interference visibility is not perfect and a fraction of the displacement pulse leaks out to the detector and may contribute to photocounts. The interference visibility $V$ is related to mode mismatch $\Delta$ by the relation $V = \sqrt{1-\Delta}$. It is now standard to achieve $ \gtrsim 95\%$ in experiment~\cite{Becerra2014}, corresponding to $\Delta \lesssim 0.1$. Additionally, I have already assumed $\alpha, \beta \in \mathbb{R}$. This is fine for PPM as the phase is not used for information encoding. Physically, however, it requires phase synchronization between the signal and displacement modes which constitutes a technical challenge---nevertheless, the degree of phase mismatch can be modeled with the $\Delta$ parameter as well~\cite{Chen2012}.

The original displacement algorithm devised by Dolinar has since been referred to as conditional pulse nulling (CPN)~\cite{Dolinar1983}, where the word \emph{nulling} entails displacements by $\beta = \alpha$. The decision strategy behind CPN is illustrated with an exemplary decision tree depicted in~\figref{fig:fig_ppm_rec}d). The idea is to null every slot and measure it with an SPD until a no-click is observed at some slot $k$, signifying $x = k$ with high probability. Then, a switchover occurs and the rest of the slots are detected directly. If no clicks are observed after the switchover, $k$ is given as the final estimate; if there are clicks detected, the output of the CPN is the output of DD in those last $M-k$ slots; if no switchover occurs, the output is random (or an erasure is announced). Proposed modifications to the CPN that further decrease its error probabilities, albeit slightly, include inexact nulling $\beta \neq \alpha$ or additional squeezing of the incoming light before the measurement~\cite{Guha2011theo,Guha2011theo2}. The error probability (allowing for inexact nulling) reads
\begin{IEEEeqnarray}{rCl}
\label{eq:PcCPNrecursive}
P_{e,M}^{\mathrm{CPN}} &=& \frac{1}{M \bar{q}_0  (\bar{q}_\beta - q_0)} \left[ (\bar{p}_0 - M \bar{q}_0) (q_0 - \bar{q}_\beta) \right. \nonumber \\
 && \left. + \; \bar{q}_0 \bar{q}_\beta^{M-1} (\bar{p}_\beta q_0 - p_0 \bar{q}_\beta) + q_0^M  (p_\beta \bar{q}_0 - \bar{p}_0 q_\beta) \right]\!,
\IEEEeqnarraynumspace
\end{IEEEeqnarray}
where $\beta$ is the constant displacement applied in the nulling rounds which can be optimized to obtain the minimal possible error probability in~\eqref{eq:PcCPNrecursive}. The slightly more involved derivation is again delegated to the Supplementary Material.

\subsection{Adaptive receivers}
Last but not least, more recently novel adaptive architectures have been put forward that optimize the detection strategy numerically by efficient means rather than brute force---either with dynamic programming~\cite{Pozza2014} or reinforcement learning~\cite{Cui2022,Bilkis2020}. Note that displacement decision tree optimization is essentially the problem of finding an optimal policy in a Markov decision process, for which dynamic programming and reinforcement learning are the customary approaches~\cite{Bellman1957}. It is worthwile to consider them separately from CPN and its modifications, as their displacement algorithm is not prescribed, but the sequence of optimal displacement amplitudes can be tailored to the observed channel behaviour. Specifically, those algorithms construct decision trees like the one in~\figref{fig:fig_ppm_rec}b) which are then traversed by the receiver according to the observed outcomes. Each node of the tree is labeled by a vector $\vec{k}$ of the so-far observed outcomes and specifies a displacement $\beta_{\vec{k}}$ to be applied in the next slot. After the measurement, the receiver proceeds along the branch corresponding to the observed measurement outcome, arrives at the next node and performs the displacement prescribed therein, so on and so forth, until it reaches the end of the tree and outputs some estimate $y_{\vec{M}}$. In principle, every displacement receiver can be described by such a decision tree and additional measurement parameters such as squeezing power can be appended to the $\beta_{\vec{k}}$ parameters to also be optimized over. The upshot of the method lies in the ability to adapt the displacement amplitude in each round to the previously observed outcomes, instead of keeping it constant for the whole duration of the measurement process, like in traditional CPN.

Unfortunately, these numerical techniques suffer from the curse of dimensionality. Already for a 32-PPM receiver with binary measurement outcomes that keeps in each node only the one-bit information whether to displace by some constant $\beta$ or not, the whole tree weighs more than a gigabyte. For 256-PPM the number of nodes comes close to the number of atoms in the observable universe. Similarly, storing the final estimates for each possible outcome sequence becomes unfeasible due to the sheer number of possibilities, and if one allows the displacements to be real numbers, the memory required to store the tree grows bigger still. On the other hand, even though adaptive methods are more efficient than brute force optimization of the displacement decision tree, the runtime of the algorithms and their sensitivity to numerical precision also increase quickly with tree depth, i.e., the modulation order. The dynamic algorithm of~\cite{Pozza2014} is shown to work for 8-PPM at the highest, whereas the largest tree constructed in~\cite{Cui2022} for quadrature amplitude modulation has depth 6 with 3 branches extending from each node. Those can surely be pushed further by upgrading the hardware with more memory and faster components, but never to $M$'s of the order of hundreds or more. Practical communication protocols are typically envisioned for $M$'s on the order of 128 or higher~\cite{PSYCHE2023} which is necessary for attaining high photon information efficiency~\cite{Banaszek2019} and bridging the gap to the Holevo limit. Therefore, optimal high-order PPM receiver architectures must necessarily operate under a modest number of rules, rather than follow a preconstructed decision tree.

\section{Results}

The outcome of this work is a new displacement receiver scheme. When applied to PPM demodulation, it offers a substantial improvement in symbol error probability over the current methods. The underlying greedy algorithm chooses the next displacement slot-by-slot to be locally optimal, that is, it maximizes the probability of a correct estimate \emph{only after} the next slot is measured, based on an efficient compression of information provided by the preceding slot measurement outcomes. This constrasts with adaptive receivers that rely on decision trees preconstructed for the whole frame duration. 

\subsection{The greedy receiver idea}

\begin{figure}[t]
\centering
\includegraphics[width=0.6\textwidth]{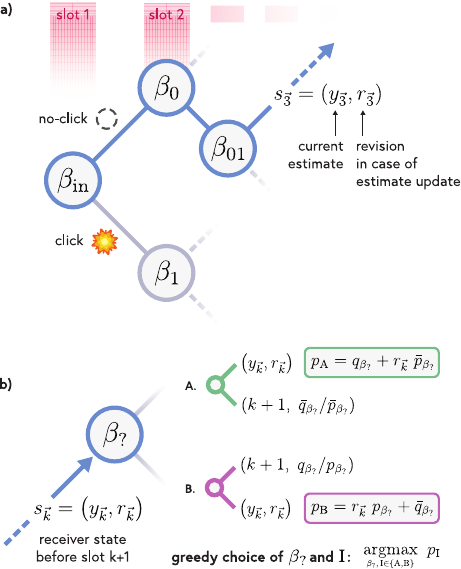}
\caption{\textbf{a) Traversing the decision tree.} After measuring $k$ slots, the greedy receiver holds in memory a temporary estimate $y_{\vec{k}}$ and a revision ratio $r_{\vec{k}}$. The latter is the quotient of a $q$-probability and a $p$-probability corresponding to the branch which it has followed after adopting the current estimate. \textbf{b) Greedy choice of the displacement and new estimates.} The receiver chooses the displacement applied in slot $k+1$ that maximizes the probability of a correct estimate after the measurement. In option A, upon recording a no-click the estimate remains unchanged from $y_{\vec{k}}$, and if a click is detected, the estimate is changed to $k+1$ along with an update of the revision ratio. In option B it is the other way around. The two options lead to different probabilities of a correct estimate, $p_{\mathrm{A}}$ and $p_{\mathrm{B}}$, both influenced by $r_{\vec{k}}$, and the receiver chooses the option leading to higher probability.}
\label{fig:fig_tree}
\end{figure}

A precise formulation of the greedy displacement algorithm is given in the Methods section, whereas here I outline the basic idea depicted in~\figref{fig:fig_tree}. The initial displacement $\beta_{\mathrm{in}}$ is given as an input to the greedy receiver, after which it assumes the initial hypothesis $y=1$ and then proceeds algorithmically. The receiver requires a small memory in which it stores the current slot hypothesis $y_{\vec{k}}$ and the so-called revision ratio $r_{\vec{k}} \in \mathbb{R}$, where $\vec{k}$ denotes the vector of the so-far observed outcomes. The revision ratio changes if after a slot measurement the receiver chooses to update its hypothesis---in this parameter it encodes the information about the future possible adjustments to the probability of a correct estimate. It can be shown that at each slot, knowing the revision ratio, the receiver can act in a way that maximizes the correct estimate probability after the next slot measurement. First of all, only two options need to be considered: \emph{A}, in which the receiver changes the estimate after recording a click and does not if no clicks are observed, and \emph{B}, in which the estimate is changed after a no-click and remains unchanged if the detector clicks. For each of the options, an optimal displacement leading to the highest correct estimate probability can be easily found numerically since it constitutes a one-parameter optimization problem. The ``greedy choice'' is then to choose option \emph{A} or \emph{B}, whichever leads to higher correct estimate probability, and apply the corresponding optimal displacement in the next slot. The receiver continues doing so until the last slot, after which the current estimate is given as the final one.

Because the greedy choice relies only on the value of the revision ratio and, importantly, in the same manner in each slot, the optimal \emph{A}/\emph{B} option and the corresponding optimal displacement can be found ahead of time for a range of $r_{\vec{k}}$ values and stored in a lookup table, in a way similar to the reinforcement-learning-based experimental setup of Ref.~\cite{Cui2022}. The optimal initial displacement $\beta_{\mathrm{in}}$ leading to the lowest output error probability can be found ahead of time by simulating the detection for a range of $\beta_{\mathrm{in}}$ values. Such a simulation is not resource-intensive even for large $M$ (for instance, I have performed it for 1024-PPM, cf.~\secref{sec:asympt}) because its runtime grows only linearly with modulation order. 

In what follows, I show how at low PPM orders for which the numerical optimization of the whole decision tree can be carried out, the greedy receiver achieves comparable results. Next, in the photon-starved regime and low noise conditions it is shown to approach the Helstrom bound with an exponential improvement in scaling compared to CPN. Because of its simplicity, the greedy receiver can be applied to arbitrarily high modulation orders as well, at which it exhibits performance analogous to low orders. Furthermore, it is shown to outperform both DD and CPN in the whole signal power spectrum, with a surprising order-of-magnitude improvement in error probability in the limit of strong pulses under noisy conditions. An exemplary application to real PPM communication scenarios concludes the Results.

\subsection{Comparison of the greedy receiver with numerical optimization}

\begin{figure}[t]
\centering
\includegraphics[width=0.6\textwidth]{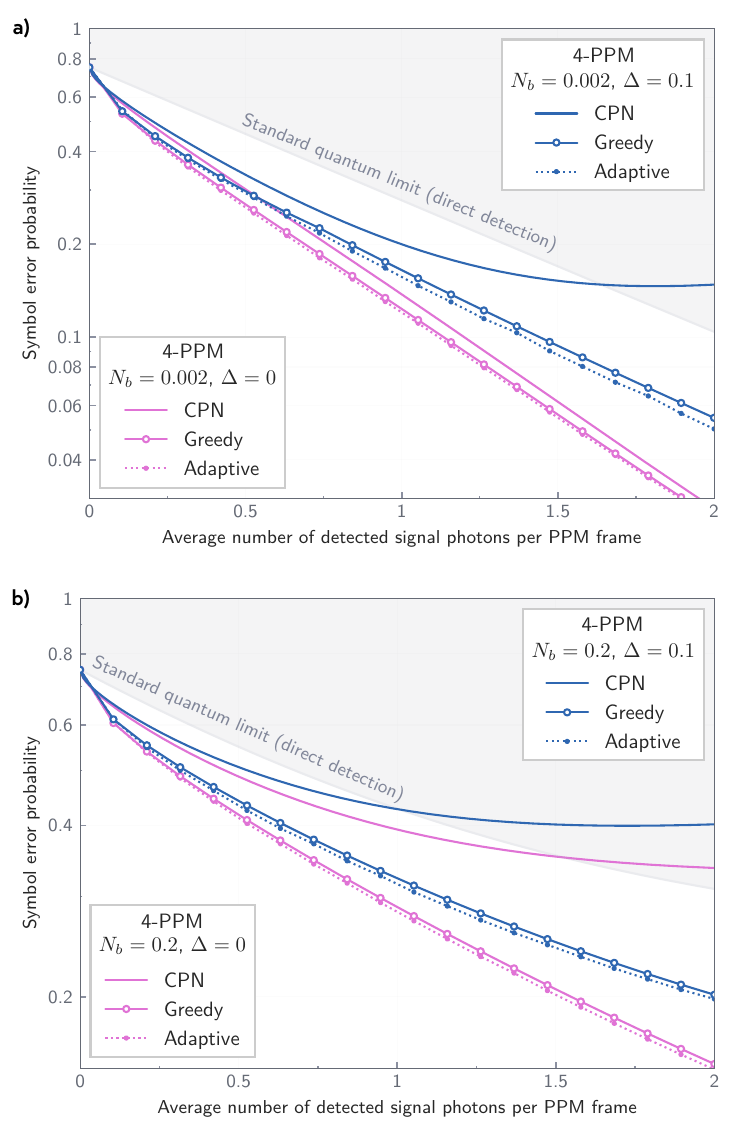}
\caption{\textbf{Receiver error probability comparison at low PPM orders for two levels of additive noise $N_b$.} Both figures depict error probabilities achievable by different PPM receivers: conditional pulse nulling (CPN), the greedy receiver and the numerically optimal adaptive receiver that can be established at low PPM orders. Additionally, the standard quantum limit of direct detection is depicted as a grey region, values below which signify an advantage over conventional detection. Within the figures, pink lines correspond to zero mode mismatch $\Delta = 0$ (perfect visibility $V = 1$) and blue lines $\Delta = 0.1$ ($V \approx 95\%$). In all cases considered, the greedy receiver closely follows the optimal error probabilities, whereas the CPN performance deteriorates quickly with increased noise and mode mismatch.}
\label{fig:fig_numcomp}
\end{figure}

For low PPM orders it is possible to numerically optimize the entire displacement decision tree, i.e., find parameters $\beta_{\vec{k}}$ and $y_{\vec{M}}$, like the ones in~\figref{fig:fig_ppm_rec}b), that yield the minimal possible average error probability. \figref{fig:fig_numcomp} compares the optimal error probabilities calculated by the adaptive algorithm of Ref.~\cite{Pozza2014} with CPN (already with optimized $\beta$ in~\eqref{eq:PcCPNrecursive}) and the greedy receiver, assuming 4-PPM. The boundaries of the grey regions correspond to the standard quantum limit (SQL), i.e., direct detection error probabilities given by~\eqref{eq:PeDD} and independent of mode mismatch $\Delta$---anything below those regions signifies an advantage over conventional detection. 

\figref{fig:fig_numcomp}a) depicts a low-noise scenario with $N_b = 0.002$, whereas \figref{fig:fig_numcomp}b) assumes noisy conditions with $N_b = 0.2$. In both figures, the pink curves correspond to zero $\Delta$ and the blue curves to $\Delta = 0.1$, modeling imperfect interference between the signal and displacement modes. In all cases, the greedy receiver closely tracks the numerically optimal performance of the adaptive algorithm. It universally outperforms DD and CPN, with the advantage pronounced especially strongly with increased $N_b$ and $\Delta$. In fact, both additive noise $N_b$ and mode mismatch $\Delta$ deteriorate the performance of CPN more quickly---even pushing it above the SQL---than that of the greedy receiver, whose error probability continues to follow the optimal one and thus can be seen to adapt well to worsening conditions. The seemingly constant proximity of the greedy and optimal results suggests the two may be somehow linked, perhaps by a constant factor---a feature encountered in some greedy algorithms~\cite{Cormen2022}.

\subsection{Asymptotic behaviour of the greedy receiver}
\label{sec:asympt}
\begin{figure}[t!]
\centering
\includegraphics[width=0.6\textwidth]{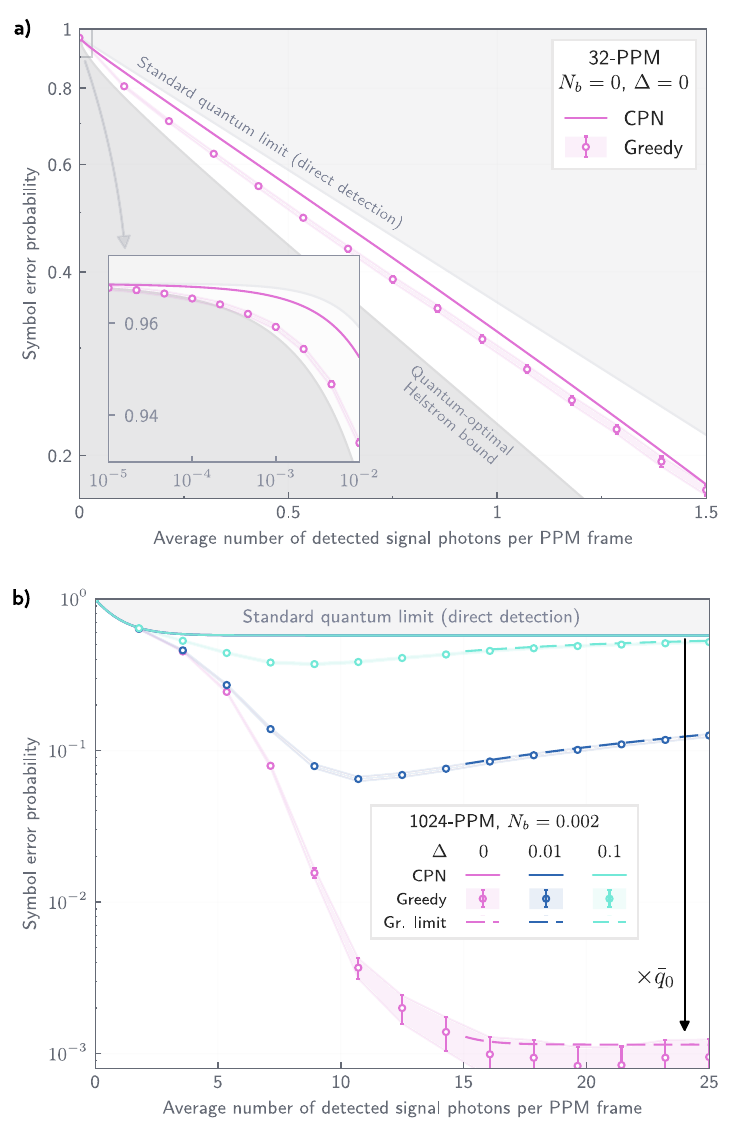}
\caption{\textbf{a) Receiver error probability comparison for noiseless conditions.} Error probabilities obtained with CPN and the greedy receiver for 32-PPM, $N_b = 0$ and $\Delta = 0$. The light gray region corresponds to conventional detection limited by the standard quantum limit, whereas the dark gray region constitutes error probabilities prohibited by quantum theory and limited by the quantum-optimal Helstrom bound. In the noiseless scenario, both the CPN and the greedy receiver achieve Helstrom-like $\sim\!\!e^{-2n}$ scaling at high photon numbers $n$, with the greedy receiver approaching CPN from below. In the photon-starved regime depicted on the inset, the greedy receiver closely follows the Helstrom bound and achieves the same scaling $\sim\!\!\sqrt{n}$, whereas CPN error probabilities scale like $\sim\!n$.  \textbf{b) Strong pulse behaviour with nonzero noise.} Receiver error probabilities for 1024-PPM assuming $N_b = 0.002$ and varying level of mode mismatch ($\Delta = 0$, $0.01$, and $0.1$, corresponding to, respectively, interference visibility $V = 100\%$, $99.5\%$, and $95\%$). A striking order-of-magnitude reduction of the greedy error probability with respect to DD\&CPN appears at high pulse energies. At this point, the other receivers reach a dark-count floor, while the greedy receiver does so only at significantly higher energies and at a lower error probability. It can be shown that for $\Delta = 0$, the greedy receiver error probability approaches a product of the SQL and empty slot click probability $\bar{q}_0$. The resulting analytical limit is shown to follow the simulated curves for nonzero $\Delta$ as well.}
\label{fig:fig_asympt}
\end{figure}

\subsubsection{Noiseless conditions}

\figref{fig:fig_asympt}a) compares the error probabilities of different PPM receivers assuming 32-PPM and ideal transmission, i.e., no additive noise $N_b = 0$ and zero mode mismatch $\Delta = 0$ in~\eqref{eq:poisson}. These assumptions allow now for a fair comparison with the quantum-optimal, noiseless Helstrom bound~\eqref{eq:Helstrom}, indicated in the Figure by a dark gray region of error probabilities prohibited by quantum theory. The standard quantum limit is again depicted as a boundary of a light gray region. 

As noted in the original work on CPN~\cite{Dolinar1983}, under noiseless conditions its error probability is \emph{near-optimal}, i.e., achieves the same large-energy scaling as the Helstrom bound. Specifically, with the average number of photons detected per frame $n \gg 0$, both the Helstrom bound~\eqref{eq:Helstrom} and the CPN error probability~\eqref{eq:PcCPNrecursive} scale like $e^{-2n}$. The greedy receiver can be seen on \figref{fig:fig_asympt}a) to approach the CPN curve with growing $n$. Indeed, in this limit the CPN error probability is an upper bound to the greedy one. This can be seen by first noting that the greedy receiver has to be initialized with some displacement $\beta_{\mathrm{in}}$, with respect to which the error probability can be minimized. Therefore, for any initial displacement the resulting greedy error probability is an upper bound to the minimal one---and if one chooses $\beta_{\mathrm{in}} = \alpha$, it can be shown (cf. Supp. Mat.) that in this limit the greedy receiver results in exactly the CPN error formula, making it an upper bound.

On the other end of the energy spectrum, the photon-starved regime $n \ll 0$ characteristic of deep-space optical communication and depicted in the inset of~\figref{fig:fig_asympt}a), the Helstrom bound scales like $\left(\frac{M-1}{M} - \gamma \sqrt{n}\right)$, where $\gamma$ is some positive proportionality factor. CPN is then no longer near-optimal, as its error probability scales linearly with $n$ like $\left(\frac{M-1}{M} - \kappa n\right)$ for some $\kappa > 0$. The greedy receiver appears to track the quantum-optimal Helstrom scaling. In this limit its analytic behaviour is not easily inferred. Therefore, I resorted to calculating the error probability formulas obtained from decision trees traversed by the greedy receiver for up to $M=12$. Indeed, in each case a $\sim\!\sqrt{n}$ scaling was observed, mimicking the quantum-optimal Helstrom bound.

\subsubsection{Strong, noisy pulses}
A striking feature of the greedy receiver is visible in~\figref{fig:fig_asympt}b) in the limit of few-photon or higher pulse energies with non-zero additive noise $N_b$, where a large gap develops between DD\&CPN and the greedy receiver error probabilities, spanning orders of magnitude. The gap diminishes with rising mode mismatch $\Delta$. Its appearance can be explained by the lack of photon number resolution, which results in DD\&CPN reaching a ``dark-count floor'', a characteristic flattening of the error probability. This arises because at this point a further increase in photon number does not influence much the pulse slot click probability as it is already practically equal to 1. The error probability is therefore limited by the amount of noise leading to clicks in empty slots. The greedy receiver has the ability to circumvent this limitation and reaches its much lower dark-count floor at higher pulse energies. 

In this regime it is also possible to analytically infer the behaviour of the greedy receiver by plugging in a chosen value of the initial displacement $\beta_{\mathrm{in}}$ and observing that the resulting error probability must then be an upper bound to the minimal possible one. From greedy receiver simulations, it appears that in the strong pulse regime, the optimal initial displacement $\beta_{\mathrm{in}}$ coincides with the pulse amplitude $\alpha$. Plugging that, it can be shown that the greedy receiver follows a specific repetitive decision tree discussed in the Supp. Mat. that contains nodes with either no displacement or displacement by $\alpha$. The resulting error probability formula is then a function of $q_0$, $p_0$, $q_\alpha$ and $p_\alpha$. Because in the strong pulse regime a click is virtually certain if a large amplitude $\alpha$ impinges on the SPD, one can take the limits $p_0 \to 0$ and $q_\alpha \to 0$, after which the resulting expression for the greedy error probability reads
\begin{IEEEeqnarray}{rCl}
\label{eq:greedy_limit}
P_{e,M}^{\mathrm{Gr. limit}} &=& \frac{\bar{p}_\alpha (-1 + M - M q_0 + q_0^M)}{M \bar{q}_0} \nonumber\\ &=& \bar{p}_\alpha \lim_{p_0 \to 0} P_{e,M}^{\mathrm{DD}} = \bar{q}_0 \lim_{p_0 \to 0} P_{e,M}^{\mathrm{DD}}.
\end{IEEEeqnarray}
The last equality follows if we additionally assume that a non-displaced pulse gives the same photodetection statistics as a displaced empty slot (this corresponds to setting $\Delta = 0$ in the photodetection statistics~\eqref{eq:poisson}). One obtains then that the greedy receiver achieves the SQL, but additionally multiplied by the probability that an empty slot clicks. The resulting formulas are plotted in~\figref{fig:fig_asympt} and found to match those obtained with greedy receiver simulation---interestingly, for $\Delta \neq 0$ as well. As an aside it can also be shown that in this regime and for $M \gg 1$ the error probabilities of DD\&CPN converge, and so the latter is in fact limited by modulation order. The gap between them and the greedy receiver is, however, independent of the modulation order.

A natural way to alleviate the dark-count floor issue is to increase the detection threshold of the DD\&CPN receivers, so that, for instance, up to two or even three photodetections in a slot are not yet counted as a click. This would effectively eliminate the erroneous clicks arising from noise but still allow for more energetic signal pulses to be detected, reducing the overall error probability. Note, however, that the error reduction in the greedy receiver is independent from the actual photodetection statistics~\eqref{eq:poisson}---in the derivation one stays at the level of $q-$ and $p-$probabilities without the need to specify them exactly. Thus, the strong pulse advantage of the greedy method over DD\&CPN persists even if the detection threshold is modified (or, in fact, for any photodetection statistics put in place of~\eqref{eq:poisson}).

\subsection{Greedy reception for deep-space communication systems}

\begin{figure}[t]
\makebox[\columnwidth][c]{\includegraphics[width=0.8\textwidth]{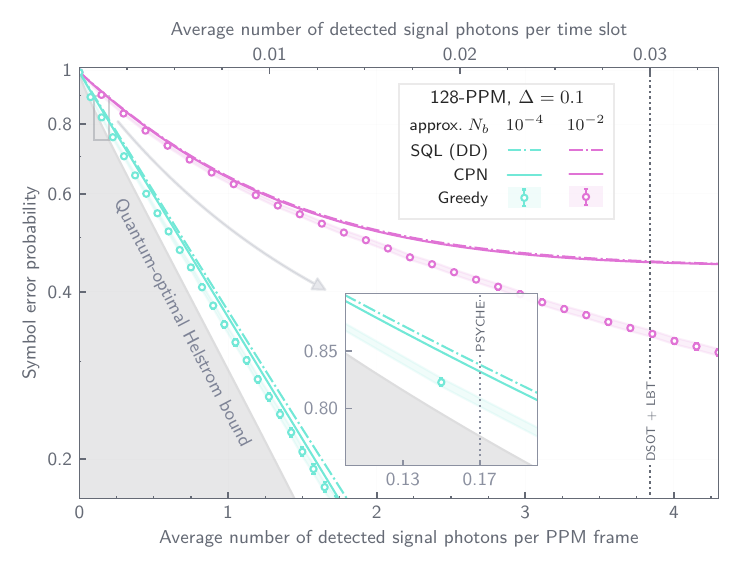}}
\caption{\textbf{Receiver error probability comparison for deep-space optical communication with 128-PPM.} The green lines correspond to the low noise conditions reported in NASA's 2023 PSYCHE mission simulations~\cite{PSYCHE2023}. The signal power (in terms of signal photons detected per frame) received from the spacecraft 1 AU away from Earth is marked on the inset. Purple lines correspond to channel conditions in~\cite{Moision2014,Jarzyna2018} that model a theorized state-of-the-art 1 AU link between the Deep Space Optical Transceiver (DSOT) and the Large Binocular Telescope (LBT) on Earth. For both cases the mode mismatch $\Delta$ has been set to 0.1, modeling a realistic value of interference visibility $V \approx 95\%$ between the signal and displacement modes.}
\label{fig:fig_real}
\end{figure}

Finally, to estimate the potential gain of greedy reception in deep-space communication scenarios, I apply it to two practical use-cases of high-order PPM: that of the recently launched PSYCHE mission~\cite{PSYCHE2023}, and that of the Deep Space Optical Transceiver concept coupled with the Large Binocular Telescope, theorized in~\cite{Moision2014}. In both cases, a conservative value of mode mismatch $\Delta = 0.1$ (interference visibility $V \approx 95\%$) has been assumed. As depicted in~\figref{fig:fig_real}, for the case of PSYCHE, operating in the photon-starved regime, the use of CPN or DD leaves a gap of approx. $0.30$ dB with respect to the Helstrom bound. The greedy receiver would allow to cut this gap in half, improving the error probability by approx. $0.15$ dB over DD\&CPN and approaching the bound within $0.15$ dB as well. For DSOT with LBT, due to stronger signals, greedy reception allows to reduce the error level by even a stronger factor when compared with DD and CPN, which additionally suffer more from the appropriately increased background noise levels.

\section{Discussion}

I have demonstrated theoretically the advantage of using a displacement receiver operating according to a greedy decision algorithm for pulse position demodulation. Experimental implementation of the receiver is well within reach of current technology---it differs from other displacement receivers~\cite{Burenkov2021} only in the displacement algorithm, which can be easily incorporated (in the form of a precomputed lookup table~\cite{Cui2022}) into the feedback loop that controls the displacements in these type of receivers. The greedy receiver exhibits improved resistance to noise and mode mismatch compared with the state-of-the-art CPN method and the conventional direct detection. Moreover, it performs comparably to the optimal displacement strategies, in the few scenarios for which the latter can be established numerically. The results raise many interesting questions. What is it about the problem studied that allows such a simple strategy to perform so well, not only in noisy conditions (for which one could expect the level of disturbance to negate the advantage of ``forward-thinking'' methods), but ideal ones as well? There exist problem structures in computer science for which greedy methods can be guaranteed to be optimal or near-optimal~\cite{Cormen2022}---perhaps such a structure could be identified in coherent state demodulation as well? The interesting next step would be to adapt the idea to other modulation formats or information-theoretic tasks based on coherent state discrimination. Preliminary results suggest the method to be applicable to the QPSK modulation format as well~\cite{burenkov_comm}.

The greedy choice idea lends itself to easy modification with regards to the number of measurement parameters and outcomes, as one still only needs to analyze a single measurement. It would therefore be interesting to incorporate, for instance, additional squeezing of the light before the measurement, like in one of the CPN modifications~\cite{Guha2011theo}, or allow for photon-number-resolution, i.e., decision trees with more than two measurement outcomes corresponding to different numbers of photons detected in each slot. One could also relax the greediness and instead of maximizing the reward after the next slot, do it for some low number of slots ahead. Any of these ideas could reduce the error probability further, possibly even bridging the gap to the Helstrom bound in the ideal scenario of~\figref{fig:fig_asympt}a). It would also be interesting to check how the greedy receiver performs in channels with time-varying characteristics as suggested in \cite{Cui2022}. Last but not least, besides error probability, the other noteworthy receiver benchmark is the achievable channel capacity~\cite{Shannon1949}. A comprehensive comparison between different receivers should then account for the possibility of erasures. This are usually announced in DD\&CPN in the event of no detections at all. A simple way to incorporate erasures into the greedy receiver would be to have it additionally store the probability of the current estimate being correct, and in the end, to only output an estimate if this probability exceeds some threshold, and otherwise declare an erasure.

\section{Methods}


\subsection{The operation of the greedy receiver}

The greedy receiver algorithm is motivated by three observations. First of all, note that any receiver is free to hold a temporary estimate $y_{\vec k} \leq k$ after measuring $k \leq M$ slots of a frame, where by the vector $\vec{k}$ I denote the history of observed outcomes. After measurement in slot $k+1$ the estimate can be kept unchanged ($y_{\vec{k+1}} = y_{\vec k}$) or it can be updated ($y_{\vec{k+1}} = k+1$). For $k=M$ the estimate $y_{\vec M}$ is the final output and the goal of a receiver is to maximize, on average, the probability of $y_{\vec M} = x$. 

Second of all, given that the receiver arrives at an estimate $y_{\vec k}$ by recording outcomes $\vec{k}$, the probability $P_{c,k}(y_{\vec{k}} = x | \vec{k})$ of the estimate being correct is a product of $p(x) = 1/M$ and a sequence of $q_\beta$ and $p_\beta$ probabilities or their barred counterparts $\bar{q}_\beta = 1 - q_\beta$, $\bar{p}_\beta = 1 - q_\beta$. Specifically, the product contains $k$ such factors, all of them $q$-like instead of the one at position $y_{\vec k}$ which is $p$-like. The $\beta$ subscripts encode the sequence of displacements applied so far and the presence of a bar indicates whether a click was observed or not in the corresponding slot. For example, assume that $y_{010} = 1$ in the general receiver tree pictured in \figref{fig:fig_ppm_rec}b). Then the probability that the receiver holds this estimate and it is correct reads $P_{c,M}(y_{010} = 1 | 010) = \frac{1}{3} p_{\beta_\mathrm{in}} \bar{q}_{\beta_0} q_{\beta_{01}}$. 

Third of all, because the expression for $P_{c,k}$ tracks the history of measurements and outcomes applied, updating it is simple. If after the measurement with displacement $\beta_?$ the receiver does not change its estimate, $P_{c,k}$ is simply multiplied by $q_{\beta_?}$ in case of a no-click or $\bar{q}_{\beta_?}$ in case of a click. On the other hand, if the receiver does update the estimate, the measurement and outcome history remain unchanged---the only required correction to $P_{c,k}$ is to change the current single $p$-like probability to a $q$-like and to multiply by $p_{\beta_?}$ or $\bar{p}_{\beta_?}$ to reflect the estimate update.

Different decisions result in different final probabilities of the estimate being correct. DD and CPN receivers make the choice according to their guiding algorithms while adaptive schemes make use of a preconstructed tree in which the choices have already been made. The idea behind the greedy receiver, depicted in~\figref{fig:fig_tree}, is to hold at all times not only the temporary estimate $y_{\vec k}$ but also the \emph{revision ratio} $r_{\vec k}$ by which one has to multiply the correct estimate probability expression to change a $p$-like probability into a $q$-like. If the temporary estimate was acquired after a no-click, the ratio is given by $r_{\vec k} = q_{\beta_{\vec k}} / p_{\beta_{\vec k}}$ where $\beta_{\vec k}$ is the displacement applied in the slot $y_{\vec k}$. Conversely, if the temporary estimate was acquired after a click, the ratio reads $r_{\vec k} = \bar{q}_{\beta_{\vec k}} / \bar{p}_{\beta_{\vec k}}$.

With the revision ratio $r_{\vec k}$ it is possible to choose $\beta_?$ such that $P_{c,k+1}$ is maximal, irrespectively of the current value of $P_{c,k}$. One needs only to compare the two options for the estimate update depicted in~\figref{fig:fig_tree}b). In option A the receiver updates the estimate in case of a no-click and keeps it unchanged in case of a click. This has the effect of multiplying $P_{c,k}$ by $p_\mathrm{A} := q_{\beta_?} + r_{\vec{k}} \bar{p}_{\beta_?}$. In option B, the estimate is updated in case of a click and kept unchanged in case of a no-click. This corresponds to multiplying $P_{c,k}$ by $p_\mathrm{B} := r_{\vec{k}} p_{\beta_?} + \bar{q}_{\beta_?}$. The revision ratio can thus be interpreted as a compression of information about the recorded outcomes which quantifies how much can be gained by an estimate update. The greedy choice is simply to displace by $\beta_?$ that maximizes $\max \{ p_\mathrm{A}, p_\mathrm{B} \}$ and assume correspondingly option A or B in the measurement.

This decision process can be repeated for every slot irrespectively of the modulation order $M$. The one-parameter optimization over $\beta_?$ should be performed beforehand for a sample of ratios $r_{\vec{k}}$ and the optimal $\beta_?$ and A/B options can be kept in a lookup table. This way in practical implementation the next displacement to apply can be found in time for the arrival of the next slot, similarly as demonstrated experimentally in~\cite{Cui2022} for adaptively learned decision trees. Note, however, that the speed of the electronics governing the displacement changes limits the minimal possible slot width, as is the case for any displacement receiver. Finally, the initial displacement $\beta_{\mathrm{in}}$ is not determined by the algorithm and should be optimized for given channel conditions by testing a range of values. 

\subsection{Simulation data}
For low PPM orders the greedy receiver error probability can be found exactly (within numerical precision) by simply generating the whole decision tree that the greedy receiver follows. This has been done in~\figref{fig:fig_numcomp} and the greedy error probabilities are plotted there as points. For larger modulation orders storing the trees becomes infeasible and because of that, in other figures the greedy receiver is simulated. For a given amplitude $\alpha$ and photodetection statistics, a lookup table is generated for a range of revision ratio values (typically for at least 1000 logarithmically spaced values of $r_{\vec{k}} \in [10^{-16}, 10^{16}]$), containing the optimal displacement to apply and greedy choice to make. Then, in one simulation round, an input symbol $x$ is drawn randomly and the slot-by-slot simulation is explicitly performed via the greedy receiver algorithm, leading to either a successful identification of the pulse position, or an error. Each point in~\figref{fig:fig_asympt} and~\figref{fig:fig_real} was simulated for 100000 rounds, and the points in the inset of~\figref{fig:fig_asympt} for 1000000 rounds. The errorbars are set to $\pm 3 \sigma$, with $\sigma$ being the sample mean standard deviation of the obtained simulation results. 

\begin{backmatter}
\bmsection{Funding}
This work was supported by the Polish Ministry of Education and Science under the ``Quantum strategies in communication through noisy optical channels'' project no. PN/01/0204/2022 carried out within the ``Pearls of Science'' program.

\bmsection{Acknowledgments}
The author is grateful to Konrad Banaszek, Ivan Burenkov, René-Jean Essiambre, Saikat Guha, Michał Jachura, Marcin Jarzyna, Jan Kołodyński, Katarzyna Michorowska, and Matteo Rosati for fruitful discussions.

\bmsection{Disclosures}
The author declares no conflicts of interest.

\bmsection{Data availability} Data underlying the results presented in this paper are not publicly available at this time but may be obtained from the author upon request.

\bmsection{Supplemental document}
See Supplement 1 for supporting content. 

\end{backmatter}

\bibliography{greedy_receiver}

\begin{thebibliography}{10}
\newcommand{\enquote}[1]{``#1''}

\bibitem{Shannon1949}
C.~E. Shannon, \enquote{Communication in the presence of noise,} {\protect\JournalTitle{Proc. IRE}} \textbf{37}, 10--21 (1949).

\bibitem{mackay2003}
D.~J. MacKay, \emph{Information Theory, Inference, and Learning Algorithms} (Cambridge University Press, Cambridge, UK, 2003), 1st ed.

\bibitem{Banaszek2020}
K.~Banaszek, L.~Kunz, M.~Jachura, and M.~Jarzyna, \enquote{Quantum limits in optical communications,} {\protect\JournalTitle{Journal of Lightwave Technology}} \textbf{38}, 2741--2754 (2020).

\bibitem{Lukanowski2021}
K.~Łukanowski and M.~Jarzyna, \enquote{Capacity of a lossy photon channel with direct detection,} {\protect\JournalTitle{IEEE Transactions on Communications}} \textbf{69}, 5059--5068 (2021).

\bibitem{Fanizza2021}
M.~Fanizza, M.~Rosati, M.~Skotiniotis, \emph{et~al.}, \enquote{Squeezing-enhanced communication without a phase reference,} {\protect\JournalTitle{{Quantum}}} \textbf{5}, 608 (2021).

\bibitem{Glauber1963}
R.~J. Glauber, \enquote{Coherent and incoherent states of the radiation field,} {\protect\JournalTitle{Physical Review}} \textbf{131}, 2766--2788 (1963).

\bibitem{Sudarshan1963}
E.~C.~G. Sudarshan, \enquote{Equivalence of semiclassical and quantum mechanical descriptions of statistical light beams,} {\protect\JournalTitle{Phys. Rev. Lett.}} \textbf{10}, 277--279 (1963).

\bibitem{Giovannetti2004}
V.~Giovannetti, S.~Guha, S.~Lloyd, \emph{et~al.}, \enquote{Classical capacity of the lossy bosonic channel: The exact solution,} {\protect\JournalTitle{Phys. Rev. Lett.}} \textbf{92}, 027902 (2004).

\bibitem{Giovannetti2014}
V.~Giovannetti, R.~Garc{\'\i}a-Patr{\'o}n, N.~J. Cerf, and A.~S. Holevo, \enquote{Ultimate classical communication rates of quantum optical channels,} {\protect\JournalTitle{Nature Photon.}} \textbf{8}, 796--800 (2014).

\bibitem{Holevo1973}
A.~S. Holevo, \enquote{Bounds for the quantity of information transmitted by a quantum communication channel,} {\protect\JournalTitle{Problems of Information Transmission}} \textbf{9}, 177--183 (1973).

\bibitem{Schumacher1997}
B.~Schumacher and M.~D. Westmoreland, \enquote{Sending classical information via noisy quantum channels,} {\protect\JournalTitle{Phys. Rev. A}} \textbf{56}, 131--138 (1997).

\bibitem{Guha2011theo2}
S.~Guha, \enquote{Structured optical receivers to attain superadditive capacity and the {H}olevo limit,} {\protect\JournalTitle{Phys. Rev. Lett.}} \textbf{106}, 240502 (2011).

\bibitem{Sasaki1998}
M.~Sasaki, K.~Kato, M.~Izutsu, and O.~Hirota, \enquote{Quantum channels showing superadditivity in classical capacity,} {\protect\JournalTitle{Phys. Rev. A}} \textbf{58}, 146--158 (1998).

\bibitem{Bilkis2020}
M.~Bilkis, M.~Rosati, R.~M. Yepes, and J.~Calsamiglia, \enquote{Real-time calibration of coherent-state receivers: Learning by trial and error,} {\protect\JournalTitle{Phys. Rev. Res.}} \textbf{2}, 033295 (2020).

\bibitem{Cormen2022}
T.~H. Cormen, C.~E. Leiserson, R.~L. Rivest, and C.~Stein, \emph{Introduction to Algorithms} (The MIT Press, 2022), 4th ed.

\bibitem{Hemmati2006}
S.~J. Dolinar, J.~Hamkins, B.~E. Moision, and V.~A. Vilnrotter, \enquote{Deep-space optical communications,} in \emph{Deep-Space Communications and Navigation Series,}  H.~Hemmati, ed. (Wiley, New York, 2006), pp. 215--289.

\bibitem{Hemmati2011}
H.~Hemmati, A.~Biswas, and I.~B. Djordjevic, \enquote{Deep-space optical communications: Future perspectives and applications,} {\protect\JournalTitle{Proceedings of the IEEE}} \textbf{99}, 2020--2039 (2011).

\bibitem{Boroson2004}
D.~M. Boroson, A.~Biswas, and B.~L. Edwards, \enquote{{MLCD: overview of NASA's Mars laser communications demonstration system},} in \emph{Free-Space Laser Communication Technologies XVI,}  vol. 5338 S.~Mecherle, C.~Y. Young, and J.~S. Stryjewski, eds., International Society for Optics and Photonics (SPIE, 2004), pp. 16 -- 28.

\bibitem{Kochman2014}
Y.~Kochman, L.~Wang, and G.~W. Wornell, \enquote{Toward photon-efficient key distribution over optical channels,} {\protect\JournalTitle{IEEE Transactions on Information Theory}} \textbf{60}, 4958--4972 (2014).

\bibitem{Jarzyna2015}
M.~Jarzyna, P.~Kuszaj, and K.~Banaszek, \enquote{Incoherent on-off keying with classical and non-classical light,} {\protect\JournalTitle{Optics Express}} \textbf{23}, 3170--3175 (2015).

\bibitem{Jarzyna2024}
M.~Jarzyna, L.~Kunz, W.~Zwoli{\'n}ski, \emph{et~al.}, \enquote{{Photon information efficiency limits in deep-space optical communications},} {\protect\JournalTitle{Optical Engineering}} \textbf{63}, 041209 (2024).

\bibitem{Helstrom1969}
C.~W. Helstrom, \enquote{Quantum detection and estimation theory,} {\protect\JournalTitle{Journal of Statistical Physics}} \textbf{1}, 231--252 (1969).

\bibitem{Cariolaro2010}
G.~Cariolaro and G.~Pierobon, \enquote{Theory of quantum pulse position modulation and related numerical problems,} {\protect\JournalTitle{IEEE Transactions on Communications}} \textbf{58}, 1213--1222 (2010).

\bibitem{Pozza2014}
N.~D. Pozza and N.~Laurenti, \enquote{Adaptive discrimination scheme for quantum pulse-position-modulation signals,} {\protect\JournalTitle{Phys. Rev. A}} \textbf{89}, 012339 (2014).

\bibitem{CoverThomas}
T.~M. Cover and J.~A. Thomas, \emph{Elements of Information Theory} (Wiley, 2006).

\bibitem{Chen2012}
J.~Chen, J.~L. Habif, Z.~Dutton, \emph{et~al.}, \enquote{Optical codeword demodulation with error rates below the standard quantum limit using a conditional nulling receiver,} {\protect\JournalTitle{Nature Photonics}} \textbf{6}, 374--379 (2012).

\bibitem{Polanskiy2010}
Y.~Polyanskiy, H.~V. Poor, and S.~Verdu, \enquote{Channel coding rate in the finite blocklength regime,} {\protect\JournalTitle{IEEE Transactions on Information Theory}} \textbf{56}, 2307--2359 (2010).

\bibitem{Jarzyna2017}
M.~Jarzyna and K.~Banaszek, \enquote{Efficiency of optimized pulse position modulation with noisy direct detection,} in \emph{2017 IEEE International Conference on Space Optical Systems and Applications (ICSOS),}  (2017), pp. 166--171.

\bibitem{Zwolinski2018}
W.~Zwoli\'{n}ski, M.~Jarzyna, and K.~Banaszek, \enquote{Range dependence of an optical pulse position modulation link in the presence of background noise,} {\protect\JournalTitle{Opt. Express}} \textbf{26}, 25827--25838 (2018).

\bibitem{Kennedy1973}
R.~S. Kennedy, \enquote{A near-optimum receiver for the binary coherent state quantum channel,} Quarterly Progress Report 108, Research Laboratory of Electronics, M.I.T. (1973).

\bibitem{Dolinar1973}
S.~J. Dolinar, \enquote{An optimum receiver for the binary coherent state quantum channel,} {\protect\JournalTitle{Quarterly Progress Report}} \textbf{111} (1973).

\bibitem{Dolinar1983}
S.~J. Dolinar, Jr., \enquote{The telecommunications and data acquisition progress report 42-72: October-december 1982,} Tech. rep., NASA, Pasadena, CA (1983).

\bibitem{Bondurant1993}
R.~S. Bondurant, \enquote{Near-quantum optimum receivers for the phase-quadrature coherent-state channel,} {\protect\JournalTitle{Opt. Lett.}} \textbf{18}, 1896--1898 (1993).

\bibitem{Burenkov2021}
I.~A. Burenkov, M.~V. Jabir, and S.~V. Polyakov, \enquote{Practical quantum-enhanced receivers for classical communication,} {\protect\JournalTitle{AVS Quantum Science}} \textbf{3}, 025301 (2021).

\bibitem{Mandel1995}
L.~Mandel and E.~Wolf, \emph{Optical Coherence and Quantum Optics} (Cambridge University Press, 1995).

\bibitem{Becerra2014}
F.~E. Becerra, J.~Fan, and A.~Migdall, \enquote{Photon number resolution enables quantum receiver for realistic coherent optical communications,} {\protect\JournalTitle{Nature Photonics}} \textbf{9}, 48–53 (2014).

\bibitem{Guha2011theo}
S.~Guha, J.~L. Habif, and M.~Takeoka, \enquote{Approaching helstrom limits to optical pulse-position demodulation using single photon detection and optical feedback,} {\protect\JournalTitle{Journal of Modern Optics}} \textbf{58}, 257--265 (2011).

\bibitem{Cui2022}
C.~Cui, W.~Horrocks, S.~Hao, \emph{et~al.}, \enquote{Quantum receiver enhanced by adaptive learning,} {\protect\JournalTitle{Light: Science \& Applications}} \textbf{11} (2022).

\bibitem{Bellman1957}
R.~Bellman, R.~Bellman, and R.~Corporation, \emph{Dynamic Programming}, Rand Corporation research study (Princeton University Press, 1957).

\bibitem{PSYCHE2023}
D.~Riel{\"a}nder, A.~D. Mira, D.~Alaluf, \emph{et~al.}, \enquote{{ESA ground infrastructure for the NASA/JPL PSYCHE Deep-Space Optical Communication demonstration},} in \emph{International Conference on Space Optics — ICSO 2022,}  vol. 12777 K.~Minoglou, N.~Karafolas, and B.~Cugny, eds., International Society for Optics and Photonics (SPIE, 2023), p. 127770E.

\bibitem{Banaszek2019}
K.~Banaszek, L.~Kunz, M.~Jarzyna, and M.~Jachura, \enquote{{Approaching the ultimate capacity limit in deep-space optical communication},} in \emph{Free-Space Laser Communications XXXI,}  vol. 10910 H.~Hemmati and D.~M. Boroson, eds., International Society for Optics and Photonics (SPIE, 2019), p. 109100A.

\bibitem{Moision2014}
B.~Moision and W.~Farr, \enquote{Range dependence of the optical communications channel,} {\protect\JournalTitle{The Interplanetary Network Progress Report}} \textbf{42-199} (2014).

\bibitem{Jarzyna2018}
M.~Jarzyna, W.~Zwoliński, M.~Jachura, and K.~Banaszek, \enquote{{Optimizing deep-space optical communication under power constraints},} in \emph{Free-Space Laser Communication and Atmospheric Propagation XXX,}  vol. 10524 H.~Hemmati and D.~M. Boroson, eds., International Society for Optics and Photonics (SPIE, 2018), p. 105240A.

\bibitem{burenkov_comm}
{Private communication with I. Burenkov} (2024).

\end{thebibliography}

\end{document}